# Multiferroics: different ways to combine magnetism and ferroelectricity [1)]

D.I. Khomskii*

*II.Physikalisches Institut, Universität zu Köln, Zülpicher Str.77, 50937 Köln, Germany*

**Abstract**

Multiferroics – materials which are simultaneously (ferro)magnetic and ferroelectric, and often also ferroelastic, attract now considerable attention, both because of the interesting physics involved and as they promise important practical applications. In this paper I give a survey of microscopic factors determining the coexistence of these properties, and discuss different possible routes to combine them in one material. In particular the role of the occupation of d-states in transition metal perovskites is discussed, possible role of spiral magnetic structures is stressed and the novel mechanism of ferroelectricity in magnetic systems due to combination of site-centred and bond-centred charge ordering is presented. Microscopic nature of multiferroic behaviour in several particular materials, including magnetite $Fe_3O_4$, is discussed.

*PACS: 77.80.-e; 64.70.Kb; 75.80.+q*

*Keywords : multiferroics ; ferroelecricity ; microscopic mechanisms*

## 1. Introduction

One of the very promising approaches to create novel materials is to combine in one material different physical properties to achieve rich functionality. The attempts to combine in one system both the (ferro)magnetic and ferroelectric (FE) properties started in 1960's, predominantly by two groups in then the Soviet Union: the group of Smolenskii in St.Petersburg (then Leningrad) [1] and by Venevtsev in Moscow [2]. Materials combining these different "ferroic" [3] properties were later on called "multiferroics" [4]. For some time this field of research was very ``quiet'' and not well known. An upsurge of interest to these problems started around 2001-2003. This is probably connected with three factors. First, the technique, especially that of preparing and studying thin films of oxides, to which most of multiferroics belong, was developed enormously. This permitted to make very good thin films of especially ferroelectric materials, and opened a possibility to use these systems e.g. for ferroelectric memory [5]. Second, several new multiferroic systems were discovered [6-9] with rather spectacular properties, in particular very strong coupling between ferroelectric and magnetic degrees of freedom. And third was probably much broader realization that with these new technical facilities and with novel materials one can think of many interesting and very promising applications, such as e.g. controlling magnetic memory by electric field or vice versa, new types of attenuators, etc.

* Corresponding author. Tel.: +49-221-470-3597; fax: +49-221-470-5178
*E-mail address*: khomskii@ph2.uni-koeln.de

From the physical point of view multiferroics present an extremely interesting class of systems and problems. These are essentially of two kinds. One is what are the microscopic conditions, and sometimes constrains, which determine the possibility to combine in one system both magnetic and ferroelectric properties. This turned out to be a quite nontrivial question, and usually, in conventional systems, these two phenomena tend to exclude one another. Why is it the case is an important and still not completely resolved issue.

Despite this apparently bad compatibility of magnetism and ferroelectricity, we now know many systems in which these properties coexist, and many more will probably be discovered in near future. The discussion of different routes to combine magnetism and ferroelectricity is the main topic of this paper As this is already a very big field in itself, some of the topics will be touched upon only very sketchy; my main aim is to give a general overview and to classify different possible ways to combine magnetism and ferroelectricity; the details can be found in the papers cited.

The second group of problems is: given the multiferroic system, what is the coupling between different degrees of freedom ? How strong is it, what are its symmetry properties, etc. These questions usually requires detailed group-symmetry analysis of a given particular system. This is an extremely important field, but I will not discuss it extensively in my paper; some examples of this approach one can find in the review articles [ 1,2,9], as well as e.g. in [11].

There is one general part in this story which has to say. The study of the coupling between magnetic and electric (dipole) degrees of freedom was initiated already long ago under the name ``magnetoelectric effect'' [12, 13]. This big field, see e.g. [14], is closely related to the new development, and also relies heavily on symmetry considerations. These aspects will be largely left out of our treatment, except may be the last section.

When considering the microscopic conditions for the coexistence of magnetism and ferroelectricity, one has to note that, whereas the microscopic nature of magnetic ordering is in principle the same in most strong magnets – it is an exchange interaction of predominantly localized magnetic moments, this is not the case with ferroelectricity. There exist many different mechanisms of FE ordering and different types of ferroelectrics. In contrast to magnetism, real microscopic mechanisms of FE are in many cases not well understood. Therefore, when discussing multiferroic systems, the main problem lies in the FE part of the story. This will also determine the structure of the present paper.

## 2. Independent magnetic and ferroelectric subsystems

The conceptually simplest situation do we meet in materials which contain separate structural units, often noncentrosymmetric, which can give rise to strong dielectric response and eventually ferroelectricity, and which simultaneously contain, somewhere else, magnetic ions. Such are for example many borates, containing $BO_3$ groups, e.g. $GdFe_3(BO_3)_4$. These materials display interesting properties, especially optical ones [15], but one should not in general expect very strong coupling between magnetic and electric degrees of freedom here, although some coupling of course should be present.

Another well-known class of such compounds, one of the first multiferroic materials, are boracites, e.g. Ni-I boracite $Ni_3B_7O_{13}I$, see e.g. [16,1,4]. This is a classical example of magnetoelectric system, on which many original ideas were tested.

Generally speaking, there may exist many other similar cases. Thus, even for hydrogen-bond ferroelectrics one can in principle think of adding magnetic ions. Again, one should not expect strong magnetic – FE coupling in all these cases, although some interesting effects can be observed.

**3. Perovskites**

As the most promising multiferroic materials one usually considers transition metal perovskites. There are a lot of magnetic materials with diverse properties among them; also most of the classical ferroelectrics, such as $BaTiO_3$ or $(PbZr)TiO_3$ (PZT), belong to this class. It is therefore not surprising that the first attempts to create multiferroic materials were mostly concentrated on this class of compounds.

However even from the first glance it becomes clear that the situation here is far from simple Thre exist hundreds of magnetic perovskites; the good collection is presented in the tables compiled by Goodenough and Longo [17]. Another, even more extensive volume in the same Landolt-Börnstein series lists hundreds of ferroelectric perovskites [18]. But the inspection of these tables shows that there is practically no overlap between these two extensive lists of materials: magnetism and FE in perovskites seem to exclude one another. Apparently the only exceptions in the stoichiometric (not mixed) perovskites are $BiFeO_3$ [11] and $BiMnO_3$ [19], and may be the recently synthesized $PbVO_3$ [20]. But even these examples in fact do not violate this general `` exclusion'' rule for perovskites: ferroelectricity in them apparently has a different source than in most of the FE of this class, such as $BaTiO_3$, see next section.

Then why this mutual exclusion ? Empirical observation is simple: all conventional FE perovskites containing transition metal (TM) ions have such ions with the formal configuration $d^0$, i.e. they have an empty d-shell (of course not all such systems are ferroelectric, the well-known example being the ``virtual'' ferroelectric $SrTiO_3$); thus this ``$d^0$-ness'' seems to be the necessary, but not sufficient conditions for FE in this class of materials. All the known FE perovskites contain TM ions with empty d-shells: $Ti^{4+}$, $Ta^{5+}$, $W^{6+}$, etc. However as soon as we have at least one, or more real d-electrons on the d-shell, such systems may be magnetic, but they are never FE.

This was probably realized by the people working in this field [1,2] already long ago, but not stressed in an apparent way and not explained. This question was raised in 1999 during the workshop on quantum magnetism at the ITP in Santa Barbara, see also [21], and was largely elaborated by N.Spaldin (N.Hill) [22]. The possible explanation is the following:

As mentioned above, the origin of magnetism is known: strong magnetism in insulators we get for partially filled inner shells (d- or f-levels). The situation with FE in perovskites is less clear. Apparently TM ions play an important role here: their off-centre shifts provide the main deriving force for FE. But why do we need empty d-shells for that? A qualitative answer (supported by ab-initio calculations) is the following: empty d-states of TM ions like $Ti^{4+}$ in $BaTiO_3$ may be used to establish strong covalency with the surrounding oxygens. And it may be favourable to shift TM ion from the centre of $O_6$ octahedra towards one (or three) oxygens, to form a strong covalent bond with this particular oxygen(s) at the expense of weakening the bonds with other oxygens, see fig.1a. The hybridization matrix element $t_{pd}$ changes by that to $t_{pd}(1 \pm gu)$, where $u$ is the distortion. In the linear approximation corresponding terms in the energy $\sim (-t^2/\Delta)$, see fig.1b, cancel ($\Delta$ is the charge-transfer gap), but in the second order in $u$ we gain some energy

$$\delta E \cong -(t_{pd}(1+gu))^2/\Delta - (t_{pd}(1-gu))^2/\Delta + + 2t_{pd}^2/\Delta = -2t_{pd}^2(gu)^2/\Delta \quad (1)$$

Also the hopping matrix element $t_{pd}$ itself in general has a nonlinear dependence on distance, such that it increases faster for shorter intersite distance; this will also increase an energy gain for off-centre distortion. If corresponding total energy gain $\sim u^2$ exceeds the energy loss due to the ordinary elastic energy $\sim Bu^2/2$, such distortion would be energetically favourable and the system would become ferroelectric.

It is clear from the fig.1b that if we have an empty d-level, only the bonding bands would be occupied (solid arrows in fig.1b), thus we only gain the electronic energy by this process. If however we have e.g. one d-electron on the corresponding d-orbital (dashed arrow in fig.1b), this electron will occupy an antibonding hybridized state, and the total energy gain will be reduced. This may be one factor suppressing the tendency to FE for magnetic ions.

But probably this is not enough to explain almost total mutual exclusion of magnetism and FE in perovskites. Thus for example one may wonder why $CaMnO_3$ or orthochromites $RCrO_3$ (R is a rare earths ion) are is not ferroelectric. In these systems we have half-filled $t_{2g}$-levels but empty $e_g$-ones. But apparently it is these latter ones which have a strong hybridization with oxygens, so that they could have produced FE by the mechanism described above. Most probably some other factors play a role here.

One such mechanism can be the following [21, 23]: Strong covalent bond with one oxygen means the formation of a singlet state, $1/\sqrt{2}(d\uparrow p\downarrow - d\downarrow p\uparrow)$ (like a usual valence bond in chemistry, e.g. in $H_2$ molecule). If however there are some other localized electrons at the TM ions, forming localized spins, for instance S=3/2 due $t_{2g}^3$ occupancy in $Cr^{3+}$ or $Mn^{4+}$, such spins would have a rather strong, $\sim J_H S$, Hund's rule exchange with the $e_g$-electrons participating in the bond formation (typically $J_H \sim 0.8$-$0.9$ eV for 3d transition metals).
This interaction ``does not like'' a singlet state and would act as a ``pair-breaker'' – much the same as the pair-breaking by magnetic impurities of Cooper pairs in singlet superconductors.

How important is this factor, is difficult to judge a-priori. Thus to check it we did the following [23]: First by using the well-established LDA+U scheme the ground state energy of $CaMnO_3$ was calculated as a function of the shift of $Mn^{4+}$ from the centre of $O_6$ octahedra towards one of the oxygens. In full calculation the energy increased with such a shift, fig 2a. This means that the centrosymmetric, i.e. non-FE structure of $CaMnO_3$ is indeed stable. Then we repeated the same calculations, artificially *switching off* the Hund's rule exchange, i.e. putting $J_H=0$, thus removing the ``pair-breaking'' effect discussed above. And in this case the energy turned out to decrease with the distortion, fig.2b. This would signal that without this pair-breaking by localized spins $CaMnO_3$ would have been unstable with respect to FE.

These calculations have been made in LMTO scheme; they have to be repeated using the better (full potential) code. But at least these first results show that indeed the breaking of a singlet valence bond by the localized spins may be instrumental in suppressing FE in magnetic materials and can at least partially explain mutual exclusion of these two types of ordering in perovskites.

What are then the ways out ? The first route was taken by the Russian groups [1,2]. They proposed to make *mixed* systems, containing both the magnetic ions and the FE-active TM ions with the $d^0$ configurations. Each of the components then can do what they ``like'': magnetic ions give some magnetic ordering, and FE-active ions make a system FE. Indeed, many such combinations of the type $AB_{1-x}B'_xO_3$ have been found, for example $PbFe_{1/2}^{3+}Nb_{1/2}^{5+}O_3$ or $PbFe_{2/3}^{3+}W_{1/3}^{6+}O_3$ [1,47] (see however next section!). Some of them have an ordered arrangement of B,B' ions, in the other they are disordered. In some of such systems transition temperatures are rather high, e.g. in $PbFe_{1/2}^{3+}Nb_{1/2}^{5+}O_3$ $T_{FE}$=387 K and $T_N$=134 K. However the coupling between FE and magnetic subsystems in them is not very strong. Still, this remains a very useful approach to the search of useful multiferroics.

Concluding this section, I want to stress once again that all the arguments presented above are applicable to systems in which TM ions are responsible for FE. Thus e.g. TM borates mentioned in the sec.2 are of different kind: FE in them apparently has nothing, or very little, to do with the TM ions present, and is due to different physical mechanisms.

**4. Bi and Pb perovskites: role of lone pairs.**

What about two apparent exceptions from the ``exclusion rule'' in perovskite family mentioned above, $BiMnO_3$ and $BiFeO_3$ ? Both these materials contain only magnetic TM ions $Fe^{3+}(d^5)$ and $Mn^{3+}(d^4)$, both are ferroelectric and simultaneously magnetic.

It seems that these two cases violate ``$d^0$''-requirement for ferroelectricity discussed in the previous section. However the more detailed treatment shows that these two cases are not really an exception from the general rule. It turns out that the main instability leading to FE in these systems is not due to TM ions as e.g. in $BaTiO_3$, but is rather driven by the A-ions, in this case Bi. $Bi^{3+}$, and also $Pb^{2+}$, are known to have the so called *lone pairs* – two valence electrons which could have participated in chemical bonds using (sp)-hybridized states (usually $sp^2$ or $sp^3$), but which in these systems do not participate in such bonds. From the phenomenological point of view this gives high polarizability of corresponding ions, which in classical theory of FE is believed to lead, or at least strongly enhance, the instability towards FE. From the microscopic point of view we can simply say that the particular orientation of these lone pairs, or dangling bonds, may create local dipoles, which finally can order in a FE or anti-FE fashion.

This qualitative picture is supported by the real ab-initio calculations [24]. These calculations have shown that indeed it is predominantly Bi lone pairs which are responsible for FE in $BiMnO3$ and $BiFeO_3$. Magnetic ordering in these systems occurs at lower temperatures than the FE one (in $BiFeO_3$ $T_{FE}$=1100 K, $T_M$=643 K; in $BiMnO_3$ $T_{FE}$=760 K, $T_M$=105 K). Coupling between these two order parameters leads to very interesting effects, which however are out of scope of this paper (see e.g. [11]).

There exist other Bi-containing ferroelectrics, some of which can be simultaneously magnetic. Such are for example the co called Aurivillius phases (layered materials containing $Bi_2O_2$ layers alternating with perovskite-type layers which can contain magnetic ions), see e.g. [36]. The coupling between magnetism and FE in them is practically not studied yet.

**5. Hexagonal manganites**

There exists yet another class of compounds which are often cited as violating the ``$d^0$-ness'' rule: hexagonal manganites $RMnO_3$ (R=Y or small rare earths). Sometimes one calls them hexagonal perovskites, although in fact it is a misnomer: despite apparently similar formula $ABO_3$, these systems have much different crystal and electronic structure. Thus, in contrast to conventional perovskites and even to quasi-one-dimensional hexagonal perovskites like $CsNiCl_3$, in $YMnO_3$ $Mn^{3+}$ ions are located not in $O_6$ octahedra, but are in a 5-fold coordination – in the centre of the $O_5$ trigonal biprysm. Similarly R-ions, e.g. Y, are not in a 12-fold, but in a 7-fold coordination. Consequently also the crystal field level scheme of Mn ions in these compounds is different from the usual one in an octahedral coordination: instead of a triplet $t_{2g}$ and a doublet $e_g$, here d-levels are split into two doublets and an upper singlet. Consequently four d-electrons of Mn3+ occupy here two lowest-lying doublets and, in contrast to Mn3+ in octahedral coordination, there is no orbital degeneracy left, so that in these compounds Mn3+ is not a Jahn-Teller ion (this may be relevant for the very existence of this particular crystal structure in these manganites [25]).

The materials $RMnO_3$ are known to be ferroelectric with pretty high transition temperatures ~900 ÷ 1000 K and with much lower Neel temperatures: thus in $YMnO_3$ $T_{FE}$ = 950 K, $T_N$ = 77 K. The nature of FE in $YMnO_3$ remained a puzzle for a long time, until the recent study [26]. Careful structural study carried out in this paper has demonstrated that in this case the off-center shifts of $Mn^{3+}$ ions from the centre of $O_5$ trigonal biprysm is very small and definitely not instrumental in providing the mechanism of FE. Apparently the main dipole moments are formed not by Mn-O, but by Y-O pairs. But this does not imply that it is Y or other R-ions which are directly responsible for FE. It turns out that the FE in these materials has completely different nature from that of for instance $BaTiO_3$.

It is well known e.g. on the example of perovskites $ABO_3$ that for small enough A-ions there occurs tilting and rotation of $BO_6$ octahedra which helps to make a close packing of the √structure and leads to a transition from cubic to orthorombic (or sometimes rombohedral) structure (the so called $GdFeO_3$ distortion). This tendency is characterised by the tolerance factor t=$(r_A+r_O)/√2(r_B+r_O)$, where $r_{A,B,O}$ are the ionic radii of corresponding ions. For small enough values of t such distortion leads to closer packing and is favourable. Typical values of cubic – orthorombic transitions due to this mechanism are ~800 ÷ 1000 K. It seems that exactly the same phenomenon occurs in $YMnO_3$ and in similar systems: to achieve close packing, rigid $MnO_5$ trigonal biprysms also tilt. But whereas in perovskites this process does not lead to FE (although also in them there appear after $BO_6$ tilting one rather short A – O bond), in the hexagonal structure of $YMnO_3$ such tilting leads to a loss of an inversion symmetry and to FE, with dipole moments mostly formed by Y – O pairs. Thus in a sense FE in these compounds is almost an ``accidental by-product'' of the tendency to close packing. It is not surprising then that also here the corresponding structural phase transitions occur at pretty high transition temperatures ~900 ÷ 1000 K.

One somewhat disappointing conclusion follows from this picture: as the mechanisms of magnetic and FE ordering in these systems have quite different nature, generally speaking one should not expect here a very strong coupling of magnetic and FE degrees of freedom – although certain coupling is definitely present and leads to quite interesting effects [10, 26].

## 6. Site- and bond-centred ordering and ferroelectricity in magnetic systems

Yet another novel mechanism of FE in magnetic materials was suggested in the recent paper [27]. We considered in this paper the question of charge ordering in TM compounds with noninteger average valence of TM ions, such as perovskites $R_{1-x}Ca_xMnO_3$ (R- rare earth metal) around x=0.5. It is well known that there often exists in these systems a charge ordering: at x≥0.5 in $LaCaMnO_3$, and even in a much broader region for smaller rare earths, e.g. for 0.3<x<0.8 for $Pr_{1-x}Ca_xMnO_3$.

Usually one treats such charge ordering (CO) as an ordering of TM ions with different valencies, or a site-centered ordering of extra electrons or holes on a metal sublattice (analogous to Wigner crystallization). Thus for half-doped manganites, x=0.5, one uses the picture of a checkerboard CO [28], fig.3(a), with alternation of, formally, $Mn^{3+}$ and $Mn^{4+}$ ions. (One should not take this terminology too literally: there is never a full charge localization with the formation of real $Mn^{3+}$ and $Mn^{4+}$ states, usually the degree of charge disproportionation is much smaller, something like 0.1÷0.2 e, so that the actual valence states in such checkerboard CO state would rather be $Mn^{(3.5 \pm \delta)+}$, with

$\delta$ ~ 0.1÷0.2. However usually the quantum numbers of corresponding states are the same as those of $Mn^{3+}$ and $Mn^{4+}$, although the actual charge density is rather delocalized among sorrounding ions due to strong covalency).

There exists however an alternative possibility: especially on the example of quasi-1d systems we know that with partial electron occupation (partial band filling) the system may be unstable with respect to Peierls distortion, for example for one electron per site (half-filled bands) – to a dimerization. One can call it a formation of a *bond-centered ordering*, or bond-centered charge density wave (CDW): all *sites* remain equivalent, but there appears an alternation of short and long *bonds*, short bonds having higher electron density.

Similar phenomenon may in principle take place also in 2d and 3d systems, including TM oxides, although in general such bond-centered structures are less favourable in these cases. Such a bond-centered superstructure, or bond CDW, was proposed for the CO state in $Pr_{1-x}Ca_xMnO_3$ at around x~0.4 [29] on the basis of a careful structural study of a good single crystal. This state, with an electron localized on a pair of Mn ions, or on a respective Mn-Mn bonds, was called Zener polaron in [29]. According to [29] these two-site polarons order in $(PrCa)MnO_3$ as shown in fig.3(b).

Theoretical study of this question, carried out in [31], has shown that indeed at certain conditions such a bond-centered ordering may be preferable as compared to the site-centered one (triangular FE region in fig.4). However the most important result was that in this region the actual solution is not a pure site-centered or bond-centered, but is an intermediate solution with both these order parameters coexisting: the state gradually transforms from the pure site-centered CO for x=0.5 to the pure bond-centered (Zener polaron) state at the left end of this region, but everywhere inside this region there is a coexistence of both. And, whereas in pure cases of figs. 3(a) and 3(b) there is no dipole moment in the system, an intermediate solution, on one hand, has well-formed dimers, and, on the other hand, left- and right, or up- and down sides of each dimer are inequivalent and have different charges, so that each dimer has a dipole moment. These moments sum up to a total nonzero dipole moment of each $MnO_2$-plane, oriented in [110] direction, see fig.3(c). If neighbouring planes would be in phase (which seems to be experimentally the case), then the whole sample would be ferroelectric. Note also that not only Mn-Mn bonds, but also Mn sites are inequivalent in this picture, thus there is no contradiction with the results of an anomalous X-ray scattering in $(PrCa)MnO_3$ [30], which were interpreted in this paper as disproving the picture of bond-centered (Zener polaron) ordering.

I am aware of only two measurements which can support the picture of FE state in Pr-Ca manganites [31, 32]. In both of them an anomaly in the dielectric constant at the charge ordering temperature was observed, this anomaly being much sharper in [32]. Unfortunately this material has a non-negligible conductivity, which can produce spurious effects e.g. due to charge accumulation at grain boundaries, etc., which can also be sensitive to CO which conductivity of the material changes. On a positive side one may say that the actual symmetry of the samples of $(PrCa)MnO_3$ studied in [29] is very low and is actually non-centrosymmetric (although in treating their data the authors of [29] used somewhat higher pseudosymmetry).

The ferroelectricity caused by the charge ordering in magnetic systems was also observed in $LuFe_2O_4$ [33], but in this case it is caused by different physical factor: with the frustration of charge ordering in the double triangular Fe layers. With the average valence $Fe^{2.5+}$, the ordering of $Fe^{2+}$ and $Fe^{3+}$ at $T_c$=320 K leads to the formation of dipoles between these layers, with the creation of total polarization, confirmed by the measurements of pyroelectric current. The coupling of magnetic and electric subsystems in this material has not been studied yet.

Concluding this section, I should mention yet another system to which the picture developed in [27] may apply – the famous case of *magnetite* $Fe_3O_4$. It is now established that below the Verwey transition at $T_V$ ~120 K there appears in $Fe_3O_4$, besides charge ordering, also a magnetoelectric effect [34], and apparently magnetite below $T_V$ is simultaneously ferroelectric [35]. As there occurs in it a ferrimagnetic ordering at rather high temperatures ~ 500 K, magnetite is apparently a *bona fide* multiferroic material. Thus it seems that $Fe_3O_4$, besides being the first magnetic material known to the mankind and the first example of an insulator-metal transition in oxides [37], may be also the first multiferroic system.

What is the microscopic nature of FE in magnetite, is not known at present. Apparently it is connected with the Verwey transition. But the detailed type of ordering below TV is still a matter of debate. Probably the best structure proposed up to date is that of [37], with an alternation of consecutive layers of $(Fe^{2+})$–(mixed $Fe^{2+}/Fe^{3+}$)–$(Fe^{3+})$–(mixed layer). However the structure deduced is monoclinic and has centre of symmetry. Apparently the real structure of $Fe_3O_4$ below Verwey transition is lower, most probably triclinic [35]. One of the possibility to get such a phase is to use the combination of site-centered and bond-centered CO, similar to the one proposed in [27]. One such possible structure is shown in fig.4. It is a combination of bond-centered structure with one short bond per $Fe_4$-tetraheder with the (somewhat simplified ) Attfield-Radaelli site-centered CO. One sees that this structure will indeed produce a net dipole moment with the polarization in b-direction, consistent with the experiment [35].

This proposal is only one of the possibilities; the actual structure may be more complicated (thus e.g. in this structure there is no doubling of unit cell in c-direction observed in [38]; this defect can be corrected along similar lines using somewhat more complicated pattern of bond and site orderings [39]). In any case, in general this concept may be a good candidate to finally solve a long-standing puzzle of the state of magnetite below Verwey transition [40].

## 7. Ferroelectricity due to magnetic ordering

The final, and may be the most interesting possibility is the generation of FE by magnetic ordering. Keeping in mind everything said above, it seems like a heresy. However in certain sense we should be open – and actually even prepared to such possibility. I want to remind the reader about a magnetoelectric effect [12, 13]: in some systems one can actually *create an electric polarization* by applying a *magnetic field*. If an external field can do it – is not it possible that the same can happen spontaneously, due to an internal field or due to certain magnetic ordering ?

This is what apparently happens in the recently discovered multiferroic materials $RMnO_3$ (with perovskite structure; R=Tb, Gd) [6], in $RMn_2O_5$ (R- different rare earths, such as Tb, Y etc.) [7], in $Ni_3V_2O_8$ [8], and in hexaferrite [9]. In all these systems FE appears in magnetically ordered state, in certain phase: the phase with the spiral ordering in $TbMnO_3$ [41], and in similar phases in other such systems.

As the ferroelectricity in these systems appears only in certain magnetically ordered states, it is probably not surprising that the coupling between magnetic and electric subsystems in them is especially strong, and one can expect giant effects. And indeed spectacular effects were observed in these systems by application of magnetic field: change of the direction of polarization in $TbMnO_3$ [6], switching from positive to negative polarization in $TbMn_2O_5$ [7], etc. Apparently also the phenomena observed in $BiFeO_3$ [11] have much in common with these systems.

What is the detailed microscopic mechanism of generation of FE by magnetic ordering in these systems, is actually not known. But one general idea [42-44] seems to be quite plausible. It seems that in most of these systems FE appears in magnetic phases with the spiral, or helicoidal magnetic structures. Thus in $TbMnO_3$ there is no electric polarization in the phase with sinusoidal magnetic structure between ~ 40 and 30 K, but nonzero **P** appears below 30 K, when magnetic structure changes from the sinusoidal to a helicoidal one [41]. Similar is the situation also in $Ni_3V_2O_8$ [8]. Qualitatively one can understand it as follows: the very notion of spiral implies that the inversion symmetry is actually already broken in it: there may be a left-moving and right-moving spiral. This in itself shows that the system is already close to become ferroelectric. By some mechanism, most probably involving spin-orbit coupling, e.g. in the form of Dzyaloshinskii's antisymmetric exchange [45], this magnetic spiral can exert an influence on a charge and lattice subsystem, producing FE, see e.g. [46].

More detailed treatment shows that the existence of a spiral magnetic structure alone is not yet sufficient for FE: not all the spiral van lead to it. As is shown by Mostovoy [44] (see also [43]), FE can appear if the spin rotation axis **e** does not coincide with the wave vector of a spiral **Q**: the polarization **P** appears only if these two directions are different, and it is proportional to the vector product of **e** and **Q**, **P**~[**e**×**Q**].

If this scenario is correct, one may expect that there should be many more multiferroic systems of this type to be discovered: any (insulating) magnet with a helicoidal magnetic structure in which spin rotation axis does not coincide with the direction of a spiral, should develop polarization in a magnetically-ordered phase. In some cases, even if the system itself does not satisfy this condition and the vectors **e** and **Q** are parallel, one may hope to change the direction of spin rotation by appropriate magnetic field (causing spin-flop transition), so that the polarization may appear above certain critical field. Such seems to be the situation in hexaferrites [9]. And in all these cases one should expect strong coupling between magnetic and ferroelectric orderings and giant mutual effects, as observed e.g. in $TbMnO_3$ [6] and in $TbMn_2O_5$[7] (although polarization itself is usually much less than in ``good'' classical ferroelectrics like $BaTiO_3$).

Unfortunately spiral magnetic structures are more common in metals (but even in these cases one may expect interesting effects connected with the induced lattice distortion violating inversion symmetry - ``ferroelectric metals''[46]). But apparently there can exist also many insulating materials with such magnetic structures, especially in frustrated systems and in those with competing interactions. In any case, all this seems to open quite a big new field of search and, hopefully, of a discovery of new multiferroics with interesting properties – the search which can now be done ``with open eyes''.

-------------- * ---------------

At this optimistic note I can end this paper, which probably does not require any extra conclusion.

**Acknowledgements :**

I am grateful to many collegues, but especially to D.Argiriou, T.Kimura, M.V.Mostovoy and N.Nagaosa for very useful discussions. This work was supported by the Deutsche Forschungsgemeinschaft through SFB 608, by the Leverhulme Foundation, UK, and by the European program "COMEPHS".

1) This paper is based on the plenary talk given at the Moscow International Symposium on Magnetism (MISM05), Moscow, June 2005

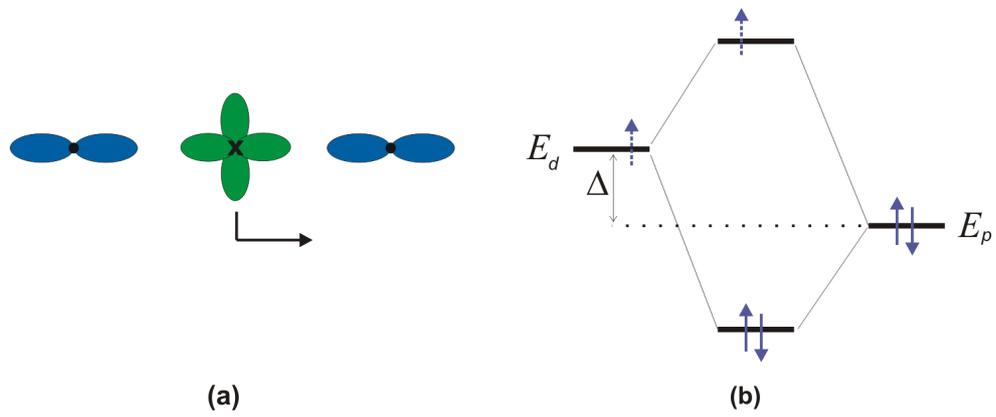

Fig.1 (color online). (a) Shifts of transition metal ion toward one of the oxygens and (b) schematic energy levels with empty d-level (solid arrows) and with partially filled d-level (dashed arrow).

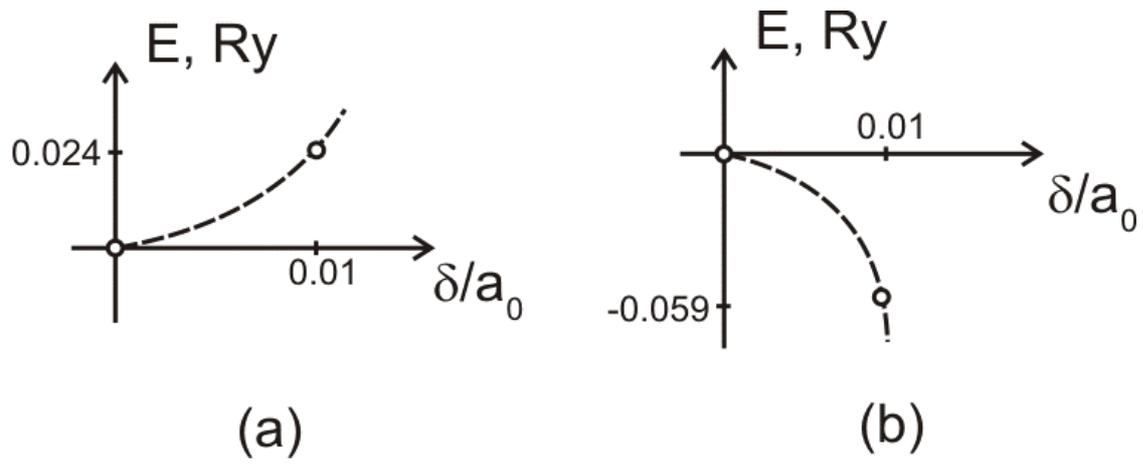

Fig.2 Change of the total energy of CaMnO$_3$ with the shift of Mn towards one of the oxygens, calculated by the LDA+U method for (a) nonzero Hund's rule coupling $J_H$=0.8 eV, and (b) for $J_H$=0 [21, 23].

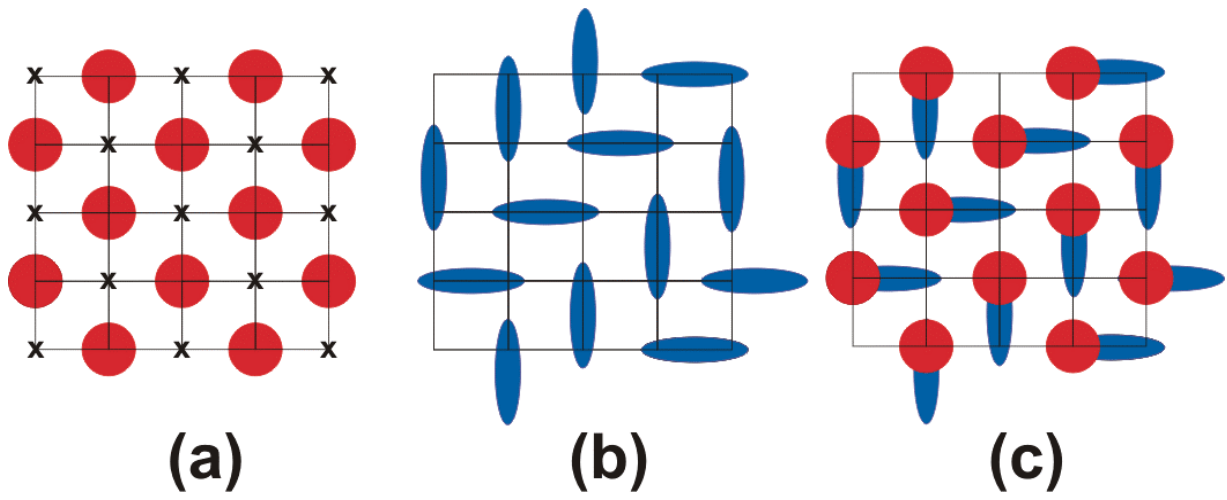

Fig.3 (color online). (a) Site-centered charge ordering in half-doped manganites like $Pr_{0.5}Ca_{0.5}MnO_3$, (b) bond-centered ordering, and (c) combined ordering, giving ferroelectricity (by [27]).

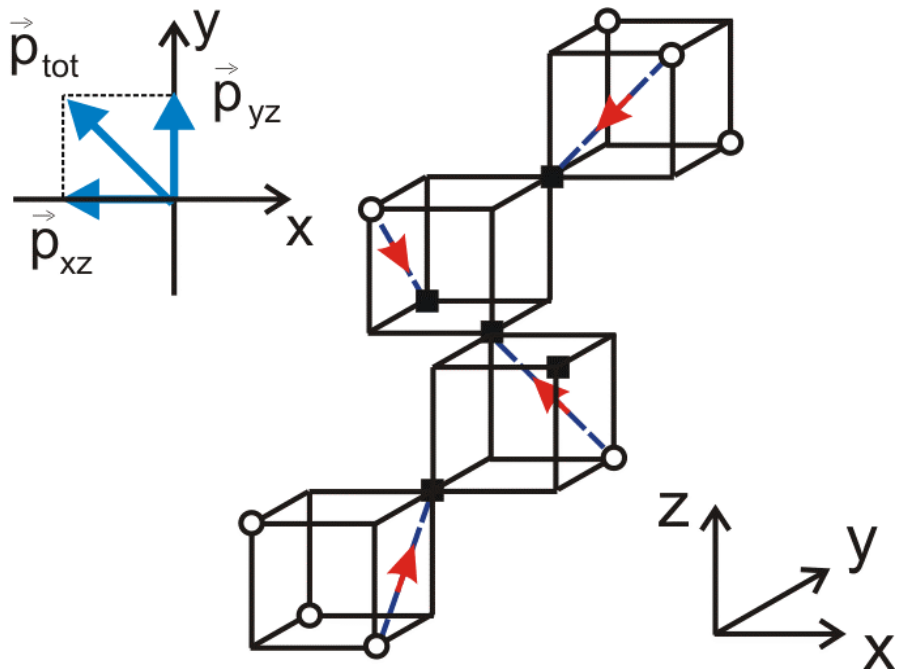

Fig.4 (color online). Possible combined bond-centered and site-centered ordering in magnetite [39] (somewhat simplified), which would give ferroelectricity below Verwey transition. Circles are $Fe^{3+}$ and squares – $Fe^{2+}$ ions. Dipole moments of each dimer are shown by arrows. Resulting spontaneous polarization **P**$_{tot}$ is shown in the inset.